\documentclass{article}

\usepackage{PRIMEarxiv}

\usepackage[utf8]{inputenc} 
\usepackage[T1]{fontenc}    
\usepackage{hyperref}       
\usepackage{url}            
\usepackage{booktabs}       
\usepackage{amsfonts}       
\usepackage{nicefrac}       
\usepackage{microtype}      
\usepackage{lipsum}
\usepackage{fancyhdr}       
\usepackage{graphicx}       
\graphicspath{{media/}}     
\usepackage{amsmath}
\usepackage{amssymb}
\usepackage{algorithm}
\usepackage{algorithmic}
\usepackage{float}
\usepackage{graphicx}
\usepackage{subcaption}

\title{UbiTouch: Towards a Universal Touch Interface
\thanks{\textit{\underline{Citation}}: 
\textbf{Authors. Title. Pages.... DOI:000000/11111.}} 
}

\author{
  Dev Shah \\
  Northwestern University, Evanston\\
  \texttt{devshah2026@u.northwestern.edu}
  }

\begin{document}
\maketitle

\begin{abstract}
Touch is one of the most intuitive ways for humans to interact with the world, and as we advance toward a ubiquitous computing environment where technology seamlessly integrates into daily life, natural interaction methods are essential. This paper introduces UbiTouch, a system leveraging thermal imaging to detect touch interactions on arbitrary surfaces. By employing a single thermal camera, UbiTouch differentiates between hovering and touch, detects multi-finger input, and completes trajectory tracking. Our approach emphasizes the use of lightweight, low-computation algorithms that maintain robust detection accuracy through innovative vision-based processing. UbiTouch aims to enable scalable, sustainable, and adaptable interaction systems for diverse applications, particularly with regards to on-human sensing.
\end{abstract}


\section{Introduction}
Touch is one of the most natural and intuitive ways for humans to interact with the world. From capacitive touchscreens on mobile phones to trackpads, touch-based interfaces have been paramount in making technology more intuitive and useful. However, these devices are often constrained to predefined surfaces, limiting their scalability and adaptability in diverse environments. As we move toward a future where computing becomes seamless and embedded in everyday life, there is a critical need for interaction systems that are portable, scalable, and capable of working across arbitrary surfaces.

Thermal imaging provides a unique opportunity to address these challenges by detecting the temperature difference between humans and typically surfaces. Unlike traditional methods, which rely on direct physical contact, thermal imaging can capture intricate hand movements in a contact-free manner, enabling interactions on any surface. This is different from regular RGB cameras due to the thermal transfer from a human touch to the surface. This capability is particularly valuable for use cases such as augmented reality, interactive displays, smart home systems, and touch-free control in sensitive environments like healthcare. 

In this paper, we introduce UbiTouch, a system that utilizes a single thermal camera to detect and classify user interactions. UbiTouch is designed to differentiate between hovering and touch, recognize multi-finger inputs, and track trajectories with high accuracy. Our approach focuses on lightweight algorithms that minimize computational complexity while maintaining robust performance, making the system practical for real-world applications for embedded devices, possibly even on-human.
\begin{figure}[!ht]
  \centering
  \includegraphics[width=0.5\textwidth]{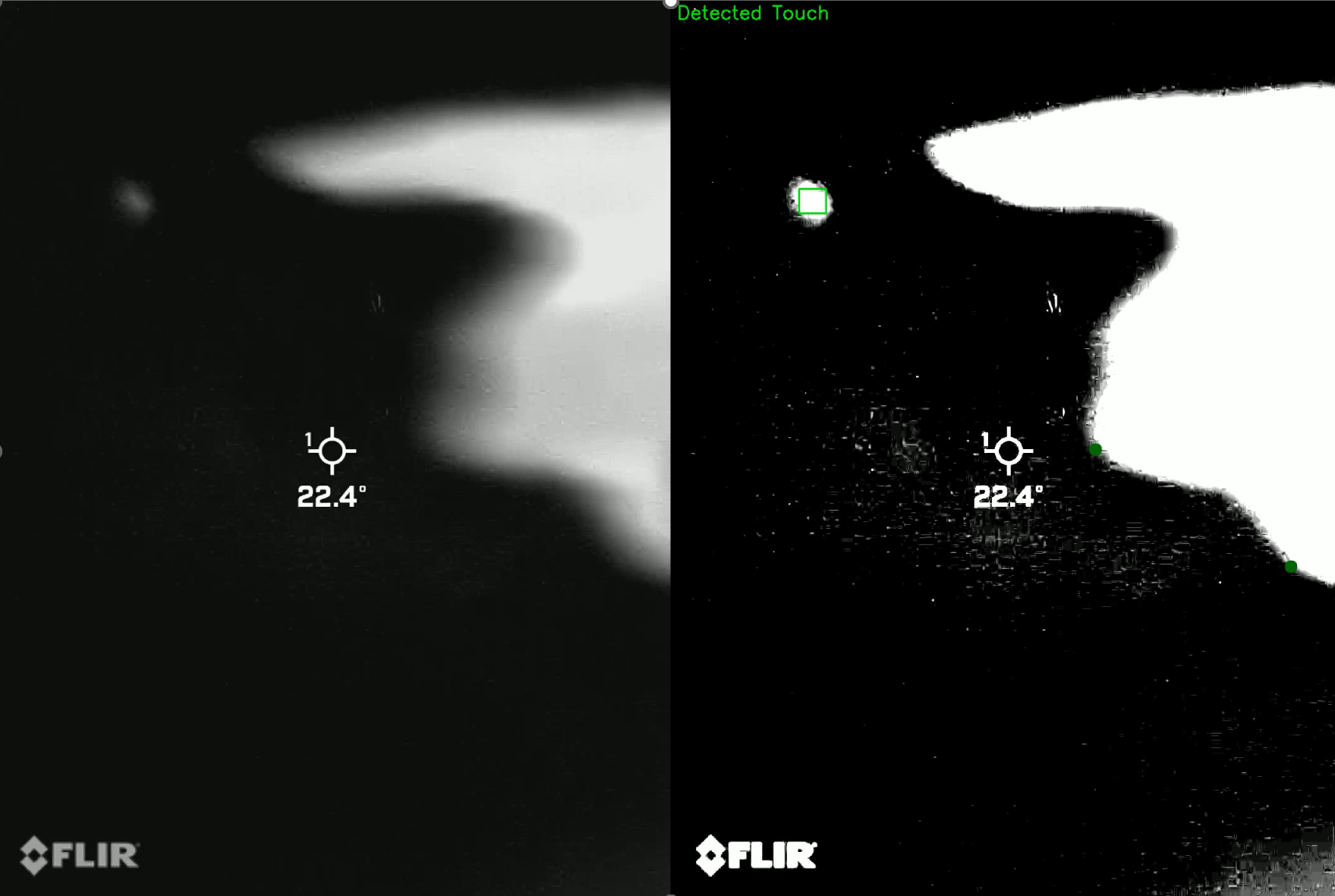}
  \caption{UbiTouch in action (image captured of a specific frame from a video). (Left) The raw thermal camera output streamed from a FLIR One Camera, showing the temperature difference between a wall and the human hand. (Right) Post processed binary image extracted along with successful detection of a touch (green box) along with fingertips (green dots). }
  \label{fig:fig1}
\end{figure}

The development of UbiTouch builds on existing research in thermal imaging and vision-based interaction systems. Through this work, we aim to advance the work being done in touch input sensing contributing to the broader goal of creating intuitive, invisible resource-constrained systems.
\section{Related Works}
UbiTouch draws upon a rich body of research across the domains of thermal imaging, vision-based systems, and multi-touch interaction interfaces. Below, we discuss relevant prior works that provide foundational insights and highlight gaps addressed by our system.

\textbf{Thermal Imaging for Interaction}
Thermal imaging has been underexplored in Human-Computer Interaction (HCI). Systems like HeatWave \cite{10.1145/1978942.1979317} and HeatGoggles \cite{10.1145/3568444.3570597} demonstrate the use of thermal cameras for interaction on arbitrary surfaces. These approaches emphasize differentiation between touch and hover events using heat transfer. However, their reliance on complex setups and limited portability restrict their application. UbiTouch enhances these contributions by focusing on a solution towards more embedded systems.
\\\\
\textbf{Feature Extraction for Touch and Hand Detection}
The use of feature extraction for robust touch and hand detection has been explored extensively. Ewerling et al. \cite{10.1145/2396636.2396663} proposed using Maximally Stable Extremal Regions (MSER) to detect fingertips and hands for multi-touch systems. Their hierarchical clustering approach allowed for robust real-time detection, even in challenging conditions like non-uniform illumination. Similarly, Jiang \cite{5235014} reviewed feature extraction techniques critical for traditional RGB computer vision systems, informing UbiTouch's algorithmic design to maintain accuracy while reducing computational overhead.
\\\\
\textbf{Multi-User Interaction and Tracking}
In multi-user environments, distinguishing between inputs from different users is crucial. Dohse et al. \cite{4455997} augmented multi-touch displays with hand tracking to improve reliability and user collaboration. By combining touch detection with hand tracking, they addressed the challenges of interference in multi-user settings. UbiTouch is built planning around similar principles, allowing for robust differentiation between individual touches, users and interactions on arbitrary surfaces.
\\\\
\textbf{Sensor Fusion}
Depth cameras have been used effectively for touch detection and tracking. Lee and Kwon \cite{s19040885} introduced a virtual touch sensor using depth imaging to detect touchpoints and correct noisy touch paths. Additionally, Dante Vision \cite{6152472} fused depth and thermal cameras to provide a more comprehensive sensing solution. UbiTouch builds on these ideas by focusing on primarily thermal imaging alone but also evaluating the benefits of simple RGB cameras to retain robust touch detection.
\\\\
\textbf{Wearable and Portable Touch Interfaces}
Wearable systems like HeatGoggles \cite{10.1145/3568444.3570597} explored using thermal imaging for touch detection on arbitrary surfaces via head-mounted devices. These systems highlight the potential for ubiquitous interaction but often target specific use cases like displays. UbiTouch expands the application space by enabling interactions across diverse surfaces, including non-display environments.
\\\\
\textbf{Active Acoustic Sensing}
The VersaTouch system \cite{10.1145/3384657.3384778} employed active acoustic sensing for multi-touch interactions on everyday surfaces. It demonstrated high accuracy in touch localization and force detection but relied on additional hardware for sensing. UbiTouch simplifies the setup by using a single thermal camera, achieving scalability and cost-effectiveness without sacrificing performance.
\\\\
\textbf{Summary}
These works provide critical insights into the design and challenges of multi-touch interaction systems. UbiTouch synthesizes these advances, focusing on scalability, computational efficiency, and adaptability for ubiquitous computing scenarios. By leveraging thermal imaging with lightweight algorithms, it bridges gaps in existing solutions and expands the possibilities for universal touch interfaces.
\section{Methodology}
UbiTouch uses a FLIR One thermal camera to implement three key features: touch detection, trajectory tracking, and jitter stabilization \footnote{A demo of UbiTouch can be seen \href{https://drive.google.com/file/d/166AZ_3Vp7iHfrEYVpGR3Tl82sZerkwLv/view?usp=sharing}{here}}. The next sections describe these in more detail. 
\subsection{Touch Detection}
\begin{figure}[!ht]
  \centering
  \includegraphics[width=0.7\textwidth]{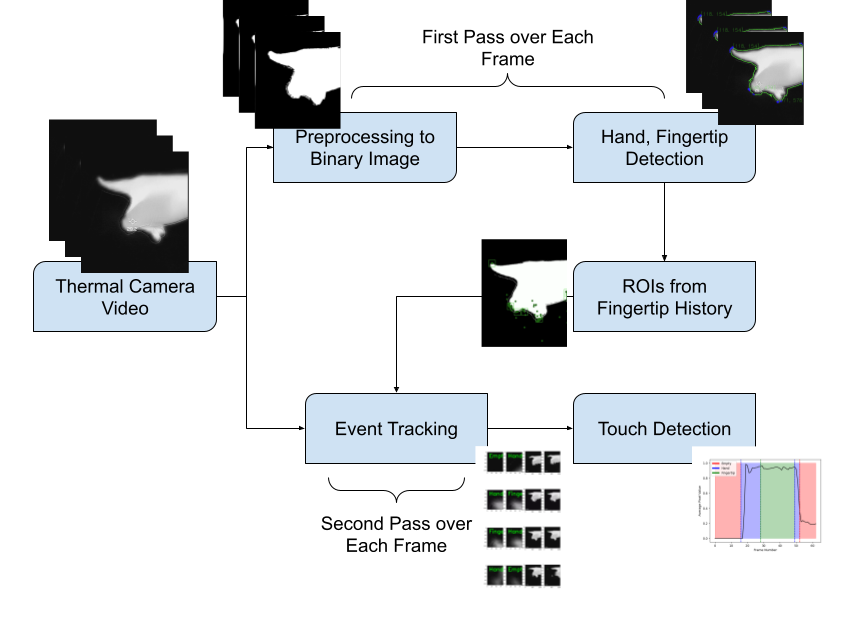}
  \caption{Theory of operation for touch detection.}
  \label{fig:flow_chart}
\end{figure}

Figure \ref{fig:flow_chart} describes the overview of operation of the touch detection algorithm. The next sections describe the steps in more detail. This section assumes that the camera is placed in a fixed position and so camera coordinates are real world coordinates. The key idea of the algorithm is that it does two passes over some input video. The first finds regions of interest (ROIs) based on where the user's fingertips have been and the second pass detects touches in these ROIs. 
\\\\
\textbf{Pre-processing:} The first step is pre-processing the video from the noisy thermal input to ideally a binary image highlighting both the human hand and any thermal transfer caused due to touch. To achieve this, following previous work, we use a grayscale version of the thermal image and apply a Gaussian blur to first remove any background noise. Next, each pixel $p_i$ is normalized by the median pixel value in the image $\bar{p}$ as follows: $\hat{p}_i = \frac{p_i - \bar{p}}{\bar{p}}$. This is done to make the temperature difference between objects in the frame relative to each other. Next, all normalized values are clipped to the range $[0, 1]$ and a binary threshold of $0.5$ is applied. To further reduce noise, some morphology operations are applied to the binary image. The resulting image now produces a very clear distinction between the background and a user's hand, given a consistently observable temperature difference.
\\\\
\begin{figure}[!ht]
  \centering
  \includegraphics[width=0.7\textwidth]{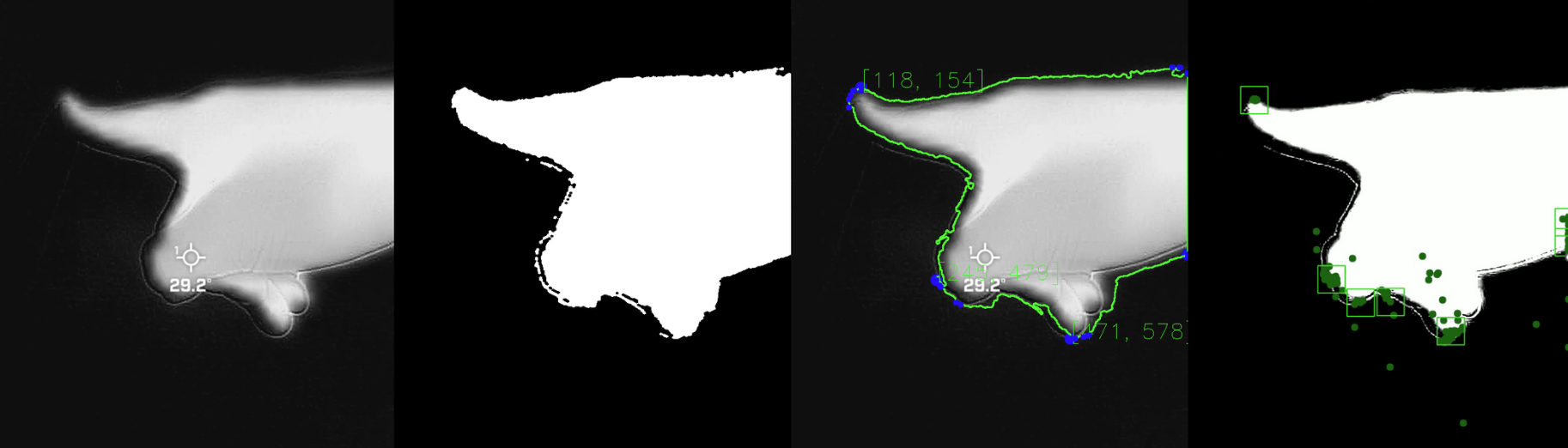}
  \caption{Fingertip and hand detection. (Left) The raw thermal camera output. (Middle left) Post processed binary image. (Middle right) Thermal image with the hand contour (green) and fingertips (blue) marked. (Right) ROIs along with histogram of fingertips marked.}
  \label{fig:img_processing}
\end{figure}
\textbf{Hand and Fingertip Detection:} Utilizing the binary image, we apply simple contour detection and choose the largest contour as an approximation to the hand of interest. We form a convex hull around this contour and calculate convexity defects. Fingertips are then identified as the end points of the convex hull that satisfy a depth threshold, filtering based on defect depth to ensure robust fingertip detection. This method provides a reliable means of detecting the hand and fingertips from thermal data as seen in figure \ref{fig:img_processing}.
\\\\
\textbf{ROI Detection:} To limit the search space and reduce false positives, we use the information about hand locations to find regions of interest (ROIs) where a possible touch event might have occurred. We simplify the problem slightly by assuming touch only occurs in regions where a fingertip has been. This is reasonable as more often than not, most intentional touch inputs involve some sort of interaction from the finger. Hence, we first go through the entire video, detecting location of fingertips at each frame using the process described above generating a scatter plot of fingerprint locations. We then implement an algorithm to find non intersecting ROIs (of a fixed size) that maximize the number of points in the scatterplot covered. The algorithm works as described in algorithm \ref{alg:roi_detection}. We now work off the assumption that touch events (if any) occur within one of these ROIs. An example of the ROIs generated by this algorithm can be seen in \ref{fig:img_processing}.
\begin{minipage}{\textwidth}
\begin{algorithm}[H]
\caption{Region of Interest (ROI) Detection Algorithm}
\label{alg:roi_detection}
\begin{algorithmic}[1]
\REQUIRE Fingertip history as a list of $(x, y)$ coordinates
\ENSURE A set of non-overlapping ROIs maximizing fingertip coverage
\STATE Mark all points as \texttt{remaining}
\WHILE{points \texttt{remaining}}
    \STATE Initialize \texttt{best\_roi}
    \FOR{each point $i$ in \texttt{remaining\_points}}
        \STATE Compute ROI bounds centered at \texttt{i} with size \texttt{roi\_size}
        \STATE Determine points within the ROI from \texttt{remaining\_points}
        \IF{This ROI contains more points that current \texttt{best\_roi}}
            \STATE Update \texttt{best\_roi}
        \ENDIF
    \ENDFOR
    \IF{\texttt{best\_roi} is \texttt{None}}
        \STATE \textbf{break}
    \ENDIF
    \STATE Add \texttt{best\_roi} to \texttt{rois}
    \STATE Mark all points within \texttt{best\_roi} as \texttt{!remaining} 
\ENDWHILE
\RETURN \texttt{rois}
\end{algorithmic}
\end{algorithm}
\end{minipage}
\\
\\\\
\textbf{Event Detection:} A touch event can be detected when there is a temperature change in the surface touched. To detect this change, we must compare the temperature of the surface before and after the touch, however as we do not have a sensor on the surface but rather a camera we cannot compare consecutive frames. This is because it is likely that the user's hand or finger touching the ROI occludes the surface immediately before and/or after the touch. However, we can instead compare the last frame before a hand/finger enters the ROI with the first frame after the hand/finger exits the ROI. This ensures that the best possible samples are compared. Thus, we need some way to extract these frames retroactively. As we have already detected the fingerprints and hands for each frame in the image, given any roi, it is simple to check if a hand or fingertip is in the roi: simply check if the point of the fingertip lies within the ROI or if the hand contour overlaps with the ROI. We can now save the frame immediately before and after any "state" change going from an empty frame to a frame with a hand/finger. 
\begin{figure}[!ht]
    \centering
    \begin{subfigure}[b]{0.55\textwidth}
        \centering
        \includegraphics[width=\textwidth]{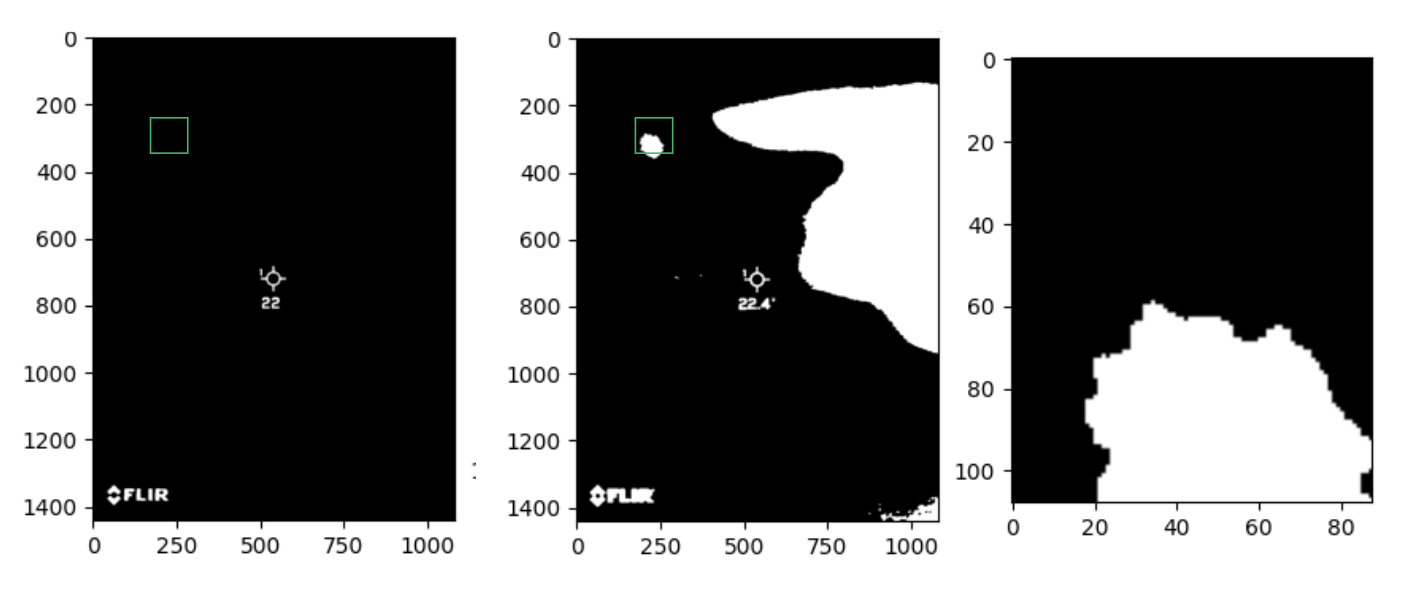}
        \caption{Change in ROI by pixel. (left) Frame immediately before hand. (middle) Frame immediately after hand exits. (right) Pixel difference of ROI post processing.}
        \label{fig:area_touch}
    \end{subfigure}
    \hfill
    \begin{subfigure}[b]{0.40\textwidth}
        \centering
        \includegraphics[width=\textwidth]{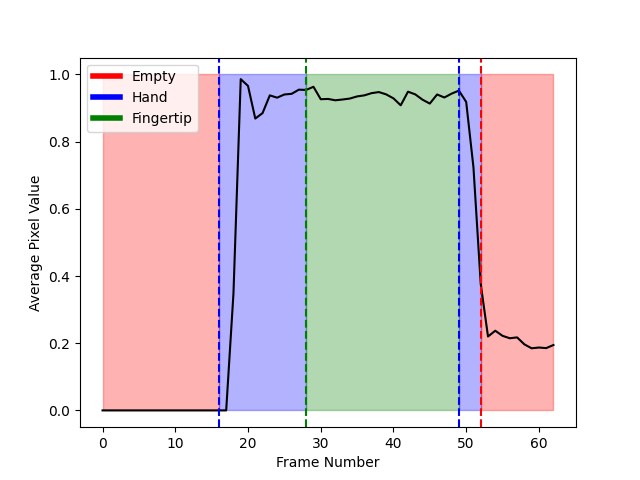}
        \caption{Average pixel value of ROI over the course of entire video. Change in average pixel value from first empty period to last empty period signifies touch.}
        \label{fig:temp_touch}
    \end{subfigure}
    \caption{Side-by-side figures showing both methods of touch detection.}
    \label{fig:side_by_side}
\end{figure}
\\\\
\textbf{Touch Detection:} Having collected the ROIs before and after change in state from/to an empty state, we can check for touch in two ways. First, we can compare the mean pixel value between consecutive empty states to see if the average temperature changed as that would be reflected in the post-processed average pixel value. On the other hand, we can also take a pixel-wise difference and after some smoothing look for large contour areas. Both of these methods work well in practice as seen in figure \ref{fig:side_by_side} and are implemented in conjunction for best results. Having successfully built the touch detection pipeline, we now discuss some desirable considerations of the algorithm. 
\\\\
\textbf{Hover vs touch:} A traditional RGB camera fails to differentiate between a touch and hover. However, as we are analyzing thermal transfer, a hover event is very different to a touch event in that even though a hover appears as an ROI, it would be rejected by both touch detectors. 
\\\\
\textbf{Mult-finger touch:} Another desirable result we achieve is multiple finger touch. As the setup allows for multiple rois, each corresponding to some (possibly) different finger(s), it ensures that any touch events across all fingers are tracked. 

\begin{figure}[!ht]
    \centering
    \begin{subfigure}[b]{0.35\textwidth}
        \centering
        \includegraphics[height=6cm]{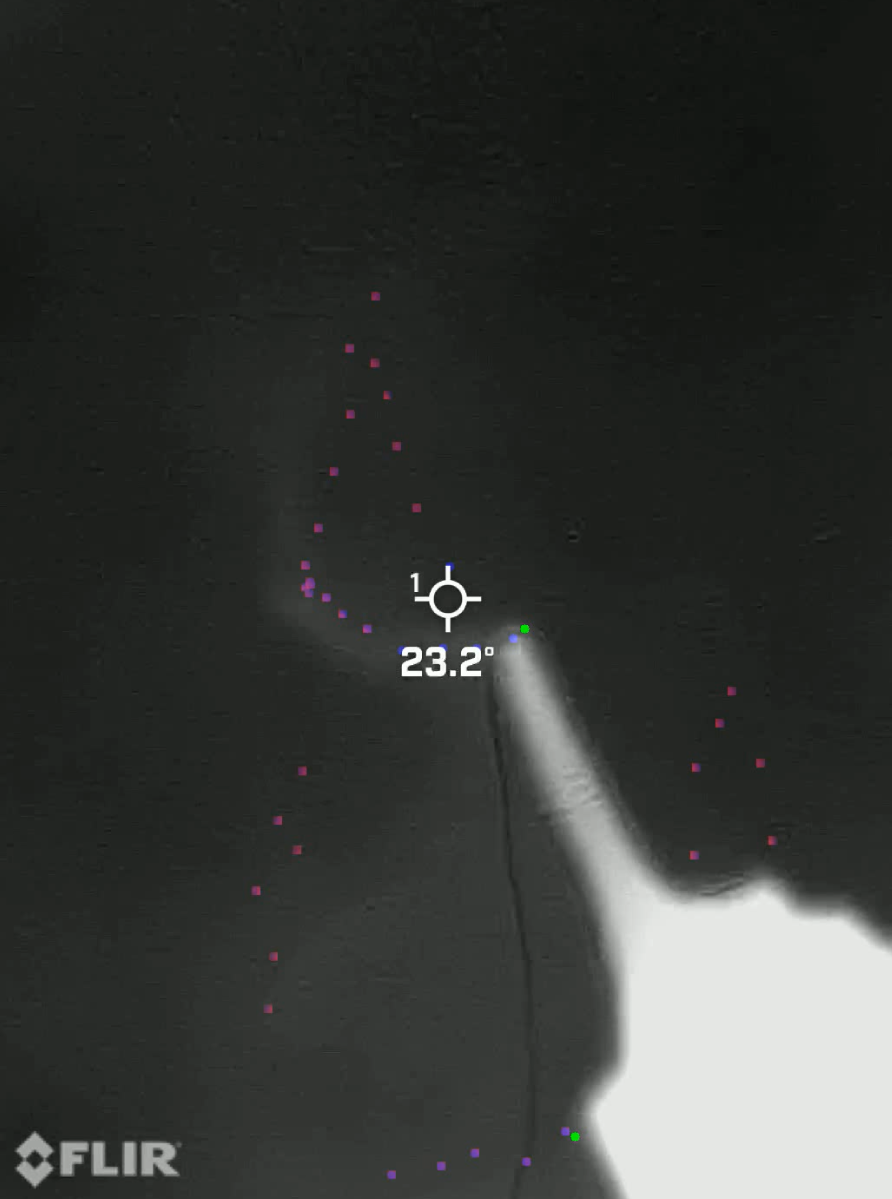}
        \caption{Heat traces reflect trajectory of finger.}
        \label{fig:traj}
    \end{subfigure}
    \hfill
    \begin{subfigure}[b]{0.6\textwidth}
        \centering
        \includegraphics[height=6cm]{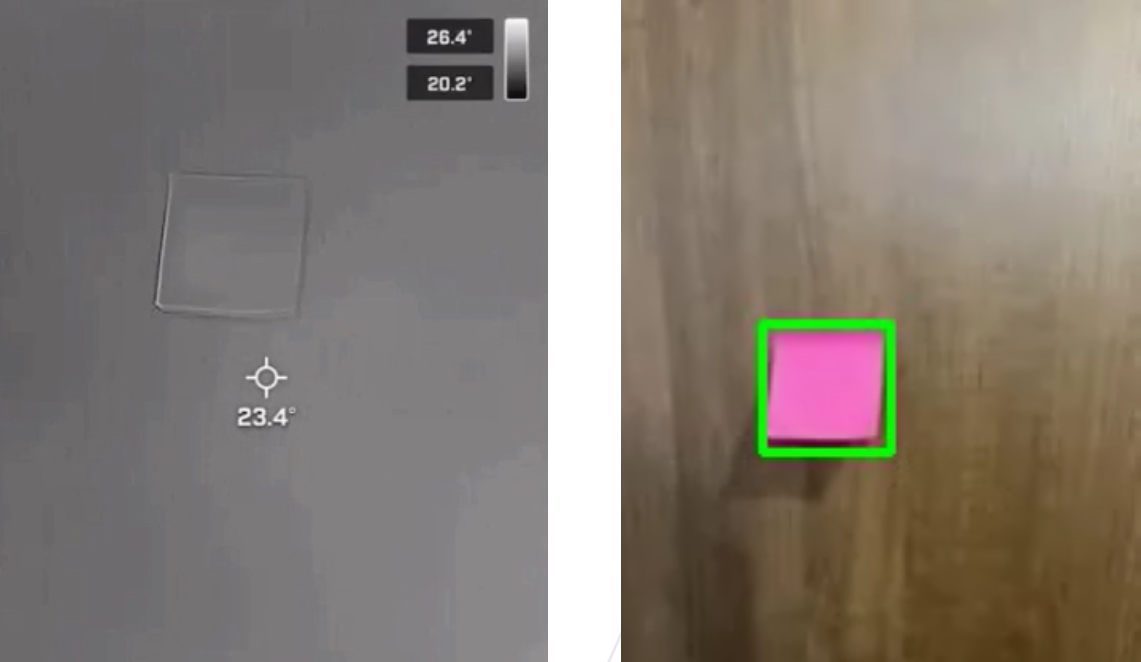}
        \caption{Jitter correction of thermal camera (left) using RGB Camera (right).}
        \label{fig:jit}
    \end{subfigure}
\end{figure}

\subsection{Trace Tracking}
The infrastructure for touch detection can now be easily extended to some other tasks. In particular, we are able to do trajectory tracking using the historical map of the fingertips. Instead of creating ROIs, we instead directly calculate temperature differences from before and after hand occluding frames. We extend this to track the temperature change between the latest empty frame and the initial pre-touch frame. By analyzing both the decay in temperature difference and the existence of a temperature difference, we can confidently reconstruct the trajectory of motion. An example of this can be seen in figure \ref{fig:traj}. 

\subsection{Jitter Reduction}
Until now we conducted experiments on a stationary camera. However, in real-life especially with on-human or wearable sensing this is not the case. A big concern due to this is the difference between camera coordinates and real-life coordinates as well as the resulting jitter. To solve this, we use sensor fusion and show some proof of concept. We place a reference object in the frame such as a sticky note in figure \ref{fig:jit} and track the camera location of the reference image from the RGB camera input. We can then construct a transformation with regards to the initial position of the reference object and apply the inverse to the thermal camera. In addition to translation, this also lets us track rotation (by comparing contours) and depth changes (by comparing area). Once the thermal camera is placed on the same coordinate system as the world, the above methods of trace tracking and touch tracking can be applied. 

\section{Proposed Evaluation Plan}
UbiTouch is a prototype and does require significant testing in the future. In order to best utilize resources, we propose a multi-pronged approach to testing. 
\\\\
\textbf{User Testing:} One definite method is a lot of user testing. Especially with multiple different users, ambient environments and use cases. This would be a good way to collect data and evaluate the prototype. Evaluation can be conducted using a custom software that allows the user to confirm or correct the algorithm's suggestion for touch, trace, etc events. 
\\\\
\textbf{Synthetic Testing:} The issue with large scale testing is the lack of thermal data for this task. Hence, another approach to testing would be to use the largely available RGB data for this task and use physics to simulate the thermal camera. This would provide interesting insights into the algorithm and allow for possibly deep learning to be implemented. Additionally, physics aware algorithms could also be implemented in the future using such an approach. 
\\\\
\textbf{Efficiency Analysis:} Another avenue of evaluation we plan to pursue is conducting a thorough survey of current touch detection solutions' algorithmic efficiency  for both thermal and traditional cameras. As a goal of UbiTouch is to work towards efficient and accurate algorithm, doing a memory and time efficiency analysis would be helpful.  
\\\\
\textbf{Preliminary Results:} We did achieve some positive preliminary results with a sample of 15 touch gestures (combination of single and multiple fingers) and 5 hover events along with 5 other negative events. Overall, a $93\%$ accuracy with the biggest cause of failure observed as false positives. 

\section{Discussion and Limitations}
UbiTouch lays a simple groundwork for low-compute 
There are two main reasons for failure for the proposed solution: (1) Incorrectly detecting hands or fingertips; and (2) incorrectly determining a touch/other interaction event. 

\textbf{Reliance on Temperature Difference:} UbiTouch is heavily reliant on the temperature difference between a human hand and the surface the hand touches. If there isn't a significant temperature difference, for example when the ambient room temperature is really high or a person's hands are really cold from being outside then the hand detection fails. Using the RGB camera to include features to detect hands would be one step towards fixing this. Another impact of temperature difference is that touch events may be detected incorrectly due to sudden changes in wall temperature. Moreover, if the temperature difference is too small, then even thermal radiations impacts it which is an issue when differentiating touch and hover.
\\\\
\textbf{Primitive hand and finger detection:} Another issue is the fact that the hand detection is quite primitive (assuming largest contour is a hand). As such, it is prone to mischaracterize hands, if say a laptop is also in frame. Another effect of this is if the hands are incorrectly detected then interaction events are immediately compromised. Here, using a more advanced yet efficient algorithm might be a valuable approach. Considering using pre-build models like mediapipe might be promising. 
\\\\
\textbf{User detection:} The current algorithm cannot detect multiple hands. However there are a lot of use cases wherein detecting different hands and users can provide interesting applications. This can be done by improving the hand detection algorithm and then running UbiTouch on each detected user's hand. 
\\\\
\textbf{Optimizing Algorithm:} The algorithm's two passes may also be parallelized in some way and made more efficient. This would go a long way towards UbiTouch's goal of efficient and accurate touch detection. 

\section{Conclusion} UbiTouch represents a significant step toward enabling scalable and efficient touch interaction systems for ubiquitous computing environments. By leveraging thermal imaging, the system overcomes limitations of traditional touch interaction solutions, enabling touch detection on arbitrary surfaces with minimal hardware requirements. The lightweight algorithms developed for touch detection, multi-finger recognition, trajectory tracking, and jitter reduction showcase the potential of thermal imaging to enhance user interactions in diverse scenarios.

While the system demonstrates promising preliminary results, several challenges remain, including reliance on temperature differences, primitive hand detection techniques, and the inability to distinguish between multiple users. Addressing these limitations through further development and integration of complementary technologies, such as RGB cameras or advanced feature extraction algorithms, will be essential to improving the system’s robustness and usability.

Future work will focus on large-scale user testing, synthetic data generation, efficiency analysis, and algorithmic optimization to refine the system and expand its applicability. Furthermore, working on the hardware of Ubitouch would also provide useful adaptation to ubiquitous computing. With continued innovation and evaluation, UbiTouch has the potential to redefine how humans interact with technology, paving the way for seamless and intuitive interfaces in everyday environments.

\section*{Acknowledgments}
This project was graciously supported by Northwestern University's Computer Science department and SPICE Lab. 

\bibliographystyle{unsrt}  
\bibliography{references}

\end{document}